\documentclass[lengthestimate,tighten,amssymb,prl,aps,twocolumn,floatfix,tightenlines,superscriptaddress,includepacs]{revtex4}
\usepackage{epsfig}
\usepackage{amsfonts}
\usepackage{amsmath}
\usepackage{graphicx}
\begin{document}

\title{Measuring a photonic qubit without destroying it

\vspace{-0.3cm}
}
\author{G. J. Pryde}
\affiliation{These authors contributed equally to this work}
\affiliation{Centre for Quantum Computer Technology, Department of Physics, University of Queensland, Brisbane 4072, Australia}
\author{J. L. O'Brien}
\affiliation{These authors contributed equally to this work}
\affiliation{Centre for Quantum Computer Technology, Department of Physics, University of Queensland, Brisbane 4072, Australia}
\author{A. G. White}
\affiliation{Centre for Quantum Computer Technology, Department of Physics, University of Queensland, Brisbane 4072, Australia}
\author{S. D. Bartlett}
\affiliation{Centre for Quantum Computer Technology, Department of Physics, University of Queensland, Brisbane 4072, Australia}
\author{T. C. Ralph}
\affiliation{Centre for Quantum Computer Technology, Department of Physics, University of Queensland, Brisbane 4072, Australia}
 \pacs{blah}

\begin{abstract}
\vspace{-0.3cm}
Measuring the polarisation of a single photon typically results in its destruction. We propose, demonstrate, and completely characterise a \emph{quantum non-demolition} (QND) scheme for realising such a measurement non-destructively. This scheme uses only linear optics and photo-detection of ancillary modes to induce a strong non-linearity at the single photon level, non-deterministically. We vary this QND measurement continuously into the weak regime, and use it to perform a non-destructive test of complementarity in quantum mechanics. Our scheme realises the most advanced general measurement of a qubit: it is non-destructive, can be made in any basis, and with arbitrary strength.
\vspace{-2cm}
\end{abstract}
\maketitle

\vspace{-2cm}
At the heart of quantum mechanics is the principle that the very act of measuring a system disturbs it. A quantum non-demolition (QND) scheme seeks to make a measurement such that this inherent \emph{back-action} feeds only into unwanted observables \cite{ca-rmp-52-341,bo-rmp-68-755}. Such a measurement should satisfy the following criteria \cite{gr-nat-396-537}: (1) The measurement outcome is correlated with the input; (2) The measurement does not alter the value of the measured observable; and (3) Repeated measurement yields the same result --- \emph{quantum state preparation} (QSP). Originally proposed for gravity wave detectors, most progress in QND has been in the continuous variable (CV) regime, involving measurement of the field quadrature of bright optical beams \cite{gr-nat-396-537}. Demonstrations at the single photon level have been limited to intra-cavity photons due to the requirement of a strong non-linearity \cite{tu-prl-75-4710,no-nat-400-239}. In addition, there has been no complete characterisation of a QND measurement due to a limited capacity to prepare input states, and thus inability to observe all the required correlations.

The importance of single-photon measurements has been highlighted by schemes for optical quantum computation that proceed via a measurement induced non-linearity \cite{kn-nat-409-46,ob-nat}. Such schemes encode quantum information in the state (\emph{eg} polarisation) of single photons --- \emph{photonic qubits}. Measurement of single photon properties is traditionally a strong, destructive measurement employing direct photo-detection. However, quantum mechanics allows general measurements \cite{nielsen} that range from strong to arbitrarily weak --- one obtains full to negligible information --- and can be non-destructive (\emph{eg} QND). Such general measurements are required \cite{sa-pra-39-694} for tests of \emph{wave--particle duality} \cite{bohm}, and other fundamental tests of quantum mechanics \cite{re-quant-ph-0310091,re-quant-ph-0310113}. They may also find application in: optical quantum computing \cite{ob-nat,ko-pra-66-063814}; quantum communication protocols \cite{bo-pra-61-050301}; tests of such protocols \cite{we-pra-47-639,na-prl-84-4733}; nested entanglement pumping \cite{du-prl-90-067901}; and quantum feedback \cite{ll-pra-62-012307}.

\begin{figure}
\begin{center}
\includegraphics*[width=7cm]{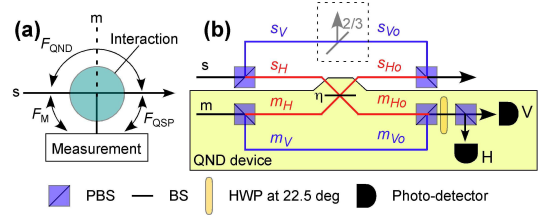}
\vspace{-0.4cm}
\caption{A QND measurement of a polarisation encoded single photon qubit. (a) A schematic of a QND measurement. After interaction with the signal $s$, measurement of a meter $m$ provides information about the signal. The performance against the requirements (1-3) can be assessed by measuring the correlations indicated by the arrows, and can be quantified by the measurement fidelity $F_M$, the QND fidelity $F_{QND}$, and the QSP fidelity $F_{QSP}$, respectively (see text for definitions). (b) A linear optics scheme to implement such a QND measurement, as explained in the text. The $\frac{2}{3}$BS is only used for weak operation.}
\label{schematic}
\vspace{-1cm}
\end{center}
\end{figure}

Here we propose, demonstrate, and completely characterise a scheme for the QND measurement of the polarisation of a free propagating single-photon qubit --- a flying qubit. This is achieved non-deterministically by using a measurement induced non-linearity. The measurement can be performed on all possible input states. Eigenstate inputs result in strong correlation with the measurement outcome; coherent superpositions exhibit ``collapse" and corresponding loss of coherence as a result of the measurement. Direct observation of all correlations demonstrates that the criteria (1-3), illustrated in Fig. \ref{schematic}(a), have been satisfied. To quantify the performance against these criteria, we introduce measures that are applicable to \emph{all} QND measurements. Finally, we show how our measurement scheme can be varied continuously from a strong measurement into the regime of a non-destructive weak measurement of polarisation. Using these weak measurements we perform a fundamental test of complementarity using ``which-path'' information without destroying the photon. Our scheme implements a measurement that is non-destructive, can be made in any basis, with arbitrary strength, and is therefore the most advanced general measurement to date.

Our scheme for QND measurement of the polarisation of a single photon in the horizontal ($H$)/vertical ($V$) basis is illustrated in Fig. \ref{schematic}(b). After interaction, a destructive measurement of the polarisation ($H$ or $V$) of an ancilliary ``meter" photon realises a QND measurement of the free-propagating ``signal" photon. The required strong optical non-linearity, which couples the signal and meter, is realised using only linear optics and photo-detection following the principles developed for optical quantum computing \cite{kn-nat-409-46,ob-nat}. As with those schemes, our QND measurement is non-deterministic: it succeeds with non-unit probability, but whenever precisely one photon is detected in the meter output, it is known to have succeeded. The key to the operation of this circuit is that the QND device makes a photon number measurement $|n=0,1\rangle$ in the $s_H$ arm of the signal interferometer: the $H$ component of the meter experiences a $\pi$ phase shift \emph{conditional} on the signal being in the state $|H\rangle_s$ (\emph{ie} in the mode $s_H$). This conditional phase shift is realised by a non-classical interference between the two photons at the $\eta$ reflectivity beam splitter (BS) and conditioned on the detection of a single meter photon.

When the signal photon is in a polarisation eigenstate, we require that the meter and signal outputs be the same state (ie $|H\rangle_s|H\rangle_m$ or $|V\rangle_s|V\rangle_m$). Consider the modes labelled in Fig. \ref{schematic}(b): the $m_V$  and $s_V$ modes are simply transmitted, while $s_{H_o}\rightarrow-\sqrt{\eta} s_H+\sqrt{1-\eta} m_H$ and $m_{H_o}\rightarrow\sqrt{1-\eta} s_H+\sqrt{\eta} m_H$. 
For the signal in the eigenstate $|V\rangle_s$, the signal and meter do not interact. We require the meter output to be $(|H\rangle_m+|V\rangle_m)/\sqrt{2}$, which is rotated by 45$^{\circ}$ to $|V\rangle_m$ by the half wave plate (HWP) set at 22.5$^{\circ}$. This is realised by preparing the meter state:
\begin{displaymath}
|D(\eta)\rangle_m=\sqrt{\tfrac{\eta}{1 + \eta}}|V\rangle_m + \sqrt{\tfrac{1}{1 + \eta}}|H\rangle_m.
\end{displaymath}
The $\sqrt{1-\eta}$ loss experienced by the $H$ component makes the $H$ and $V$ components equal and the signal-meter output state is:
\begin{displaymath}
|\phi^{V_s}_{out}\rangle=
\sqrt{\tfrac{\eta}{1+\eta}}|V\rangle_s(|V\rangle_m+|H\rangle_m)+\sqrt{\tfrac{1-\eta}{1+\eta}}|H\rangle_s|V\rangle_s,
\end{displaymath}
where the first term represents successful operation, and the second a failure mechanism corresponding to two photons in the signal output and no photons in the meter output. After rotation of the meter state by 45$^\circ$ the successful output state is $|V\rangle_s|V\rangle_m$.

When the signal is in the other eigenstate input $|H\rangle_s$ the output state is
\begin{displaymath}
\label{hsout}
|\phi^{H_s}_{out}\rangle=
\sqrt{\tfrac{1}{1+\eta}}[(1-2\eta)|H\rangle_s|H\rangle_m-\eta|H\rangle_s|V\rangle_m]+...
\end{displaymath}
where the terms not shown are ones with two photons in one of the outputs and zero in the other. We require the coefficients be equal so that the meter state is $(|H\rangle_m-|V\rangle_m)/\sqrt{2}$ (which is rotated to $|H\rangle_m$ by the HWP). This is only satisfied for $\eta=\tfrac{1}{3}$, and thus we prepare the meter in the state
\begin{displaymath}
|D'\rangle_m\equiv|D(\tfrac{1}{3})\rangle_m=\tfrac{1}{2}|V\rangle_m+\tfrac{\sqrt{3}}{2}|H\rangle_m.
\end{displaymath}

The probability of success for an arbitrary input $\gamma|H\rangle_s+\delta|V\rangle_s$ is $P=(\gamma^2+3\delta^2)/6$. The fact that $P$ is dependent on the input state must be taken into account when inferring populations from repeated measurements of identically prepared input states. It is possible to introduce the $\frac{2}{3}$ loss shown in the $s_V$ mode in Fig. \ref{schematic}(b) to make $P=\frac{1}{6}$ independent of $\delta$ and $\gamma$, as done below for weak measurement operation. Successful QND would then be signalled by the detection of a single photon in the meter output and no photon in this extra loss mode.

\begin{figure}
\begin{center}
\includegraphics*[width=7cm]{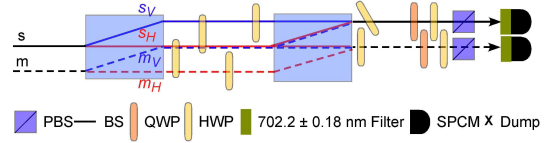}
\vspace{-0.4cm}
\caption{A schematic of the experimental setup. Pairs of photons of wavelength $\lambda=702.2$ nm were injected from the left. They were generated in a 5 mm non-linear $\beta$-barium-borate (BBO) crystal through spontaneous parametric down conversion of a $\lambda=351.1$ nm, $P\simeq300$ mW pump beam. The BBO crystal was cut for beam-like \protect\cite{ta-ol-26-843,ku-jmo-48-1997} type-II phase matching. Collection of these beam-like modes into two single mode optical fibers was optimized by focussing the pump beam to a waist at the crystal equal to the waist of the mode collected by the fibers \protect\cite{ku-pra-64-023802}. The output of each fiber was collimated and wave plates in each input beam allows preparation of the required signal $s$ and meter $m$ states. Polarisation beam displacers are used to form stable signal and meter Mach-Zehnder polarisation interferometers. Note that in contrast to the schematic of Fig. \ref{schematic}(b), the $s_V$ and $m_H$ modes interfere non-classically. In the centre region each mode passes through two HWPs. The combined effect is to rotate the polarisation of $s_H$ and $m_V$ by 90$^\circ$ and $s_V$ and $m_H$ by 125$^\circ$. This allows the polarisation modes to recombine correctly at the second beam displacer and implents the effective $\frac{1}{3}$BS. The tilted HWP corrects a systematic phase shift between the meter modes. A standard polarization analyzer --- a HWP, QWP and PBS --- is used in each output before a SPCM. A coincidence window of 5 ns was used with coincident counts of $\sim$100 s$^{-1}$, and we do not subtract accidental counts.}
\label{exp}
\vspace{-0.7cm}
\end{center}
\end{figure}

Figure \ref{exp} outlines our experimental design for realising the schematic circuit of Fig. \ref{schematic}(b). The data collected are coincident counts --- simultaneous detection of a single photon at each of the detectors --- for two reasons. First, in order to characterise the QND measurement we need to measure the polarisation of both the signal and meter photons to determine the correlation between them. By adjusting our analysers we can directly measure the probability $P_{HH}$ of the two photons being horizontally polarised, etc. Second, although in principle measurement of a \emph{single} meter photon would indicate that the measurement worked, currently available single photon counting modules (SPCMs) cannot distinguish between one and many photons and operate with moderate efficiency. Note that wave plates can be used to measure the signal in any basis.

\begin{table}[t!]
  \centering
  \caption{Experimental values for the joint probabilities for the signal and meter polarization $P_{sm}$. For the inputs $|D^-\rangle$ and $|R^-\rangle$ (not shown) results are almost identical to those for $|D^+\rangle$ and $|R^+\rangle$.}\label{prob}
\begin{tabular}{|c|c|c|c|c|}
\hline
Signal input & $|H\rangle_s$ & $|V\rangle_s$ & $|D^+\rangle$ & $|R^+\rangle$ \\
\hline
\hline
$P_{HH}$ & 0.97(3) & 0.012(3) & 0.44(3) & 0.46(3)\\
\hline
$P_{HV}$ & 0.024(3) & 0.00013(7) & 0.016(3)  & 0.022(3)\\
\hline
$P_{VH}$ & 0.007(1) & 0.18(1) & 0.10(1) & 0.104(8)\\
\hline
$P_{VV}$ & 0.0005(3) & 0.81(4) & 0.44(3) & 0.41(2)\\
\hline
\end{tabular}
\vspace{-0.4cm}
\end{table}

We prepared the signal in the eigenstates $|H\rangle_s$ and $|V\rangle_s$, and the superposition states 
\begin{eqnarray*}
&&|D^{\pm}\rangle_s\equiv\tfrac{1}{2}|V\rangle_s\pm\tfrac{\sqrt{3}}{2}|H\rangle_s,\   |R^{\pm}\rangle_s\equiv\tfrac{1}{2}|V\rangle_s\pm i\tfrac{\sqrt{3}}{2}|H\rangle_s
\end{eqnarray*}
(states which give equal probability of measuring $|H\rangle_s$ and $|V\rangle_s$), and measured the probabilities $P_{sm}=P_{HH}$, $P_{HV}$, $P_{VH}$, and $P_{VV}$ (Table \ref{prob}). The QND measurement works most successfully in the case of a $|H\rangle_s$ signal, because it requires only the splitting and recombining of the meter components. In contrast, all other measurements require both classical and non-classical interference.

To quantify the performance of a QND measurement relative to the criteria (1-3), we define new measures that can be applied to all input states. These measures each compare two probability distributions $p$ and $q$ over the measurement outcomes $i$, using the (classical) fidelity
\begin{displaymath}
F(p,q)=(\sum_i\sqrt{p_i q_i})^2.
\end{displaymath}
For photonic qubits, $i\in\{H,V\}$; also, $F=1$ for identical distributions, $F=\frac{1}{2}$ for uncorrelated distributions, and $F=0$ for anti-correlated distributions. For a QND measurement there are three relevant probability distributions: $p^{in}$ of the signal input, $p^{out}$ of the signal output, and $p^m$ of the measurement. These distributions, and hence fidelities, are functions of the signal input state. The requirements (1-3) demand \emph{correlations} between these distributions [see Fig. \ref{scematic}(a)] as follows:

(1)  The success of the measurement is quantified by the \emph{measurement fidelity} $F_M=F(p^{in},p^m)$, which measures the overlap between the signal input and measurement distributions. For signal eigenstates, we measure  $F_M(|H\rangle_s) = P_{HH} + P_{VH} = 0.97\pm0.03$ and $F_M(|V\rangle_s) = P_{VV} + P_{HV} = 0.81\pm0.04$.  For all superposition states, $|D^{\pm}\rangle_s$ and $|R^{\pm}\rangle_s$, $F_M>0.99$.

(2)  For the measurement to be \emph{non-demolition}, the signal output probabilities should be identical to those of  the input. This is characterised by the \emph{QND fidelity} $F_{QND}=F(p^{in},p^{out})$. For all signal inputs measured (eigenstates and superpositions), $F_{QND}>0.99$. 

(3) When the measurement outcome is $i$, a good QSP device gives the signal output state $|i\rangle_s$ with high probability. We denote this conditional probability $p_i^{out}|i$ and define the \emph{QSP fidelity} $F_{QSP}=\sum_i p_i^m p_i^{out}|i$, which is an average fidelity between the expected and observed conditional probability distributions. For our scheme $F_{QSP}=P_{HH}+P_{VV}$.  The average for the six inputs quantifies the performance as a QSP device for \emph{any} unknown input, and is $0.88\pm0.05$. This quantity is also known as the \emph{likelihood} $L$ \cite{wo-prd-19-473} of measuring the signal to be $H$ or $V$ given the meter outcome $H$ or $V$, respectively.

In the CV regime, $L=0$ due to the continuous spectrum of the measurement outcome. To compare directly with CV experiments, the QSP performance could also be quantified by the \emph{correlation function} \cite{qndnote1} between the signal and meter: $P_{HH} + P_{VV} - P_{HV} - P_{VH}$. This correlation is also referred to as the \emph{knowledge}; for qubits, $K = 2L -1$.  Both $K$ and $L$ are useful for characterising the weak measurements, which we now describe.

\begin{figure}[t!]
\begin{center}
\includegraphics*[width=4.5cm]{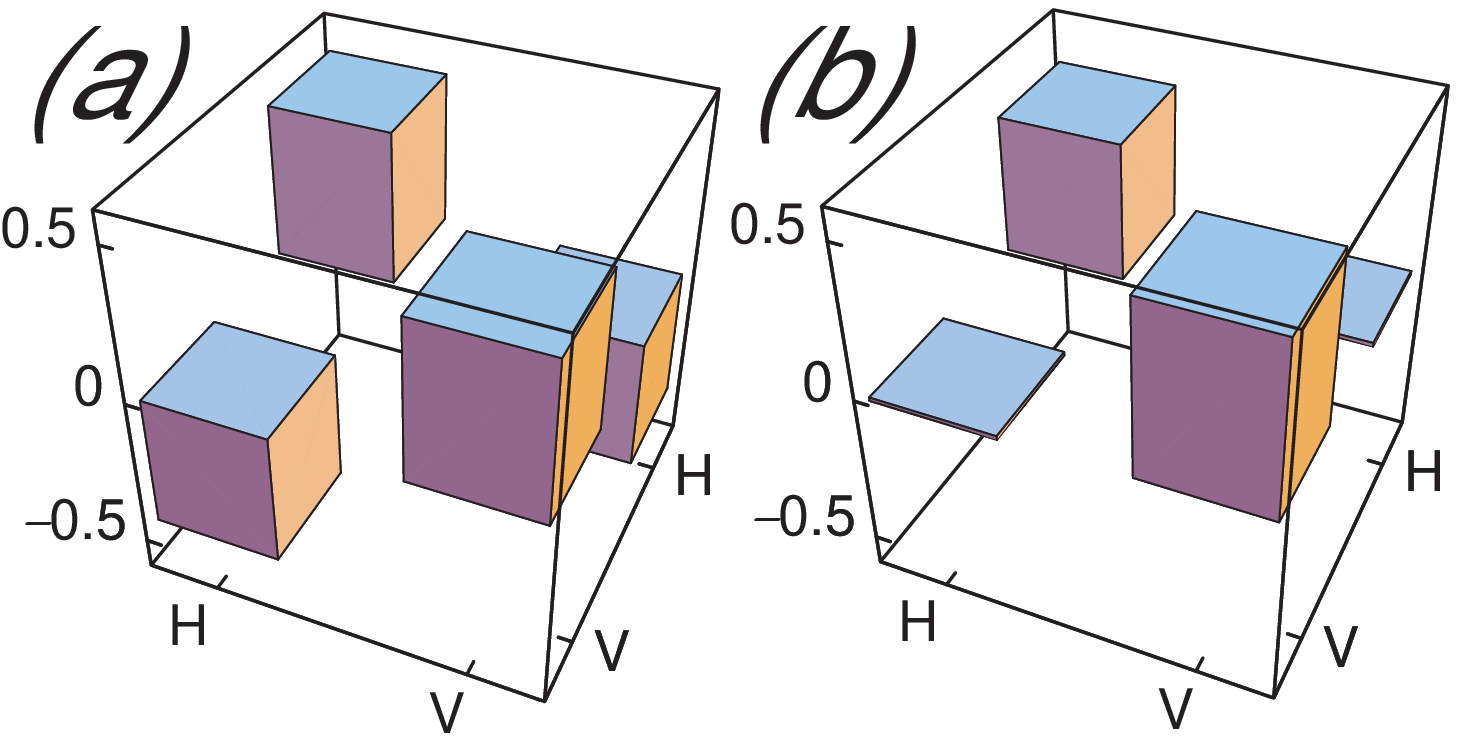}
\includegraphics*[width=6.5cm]{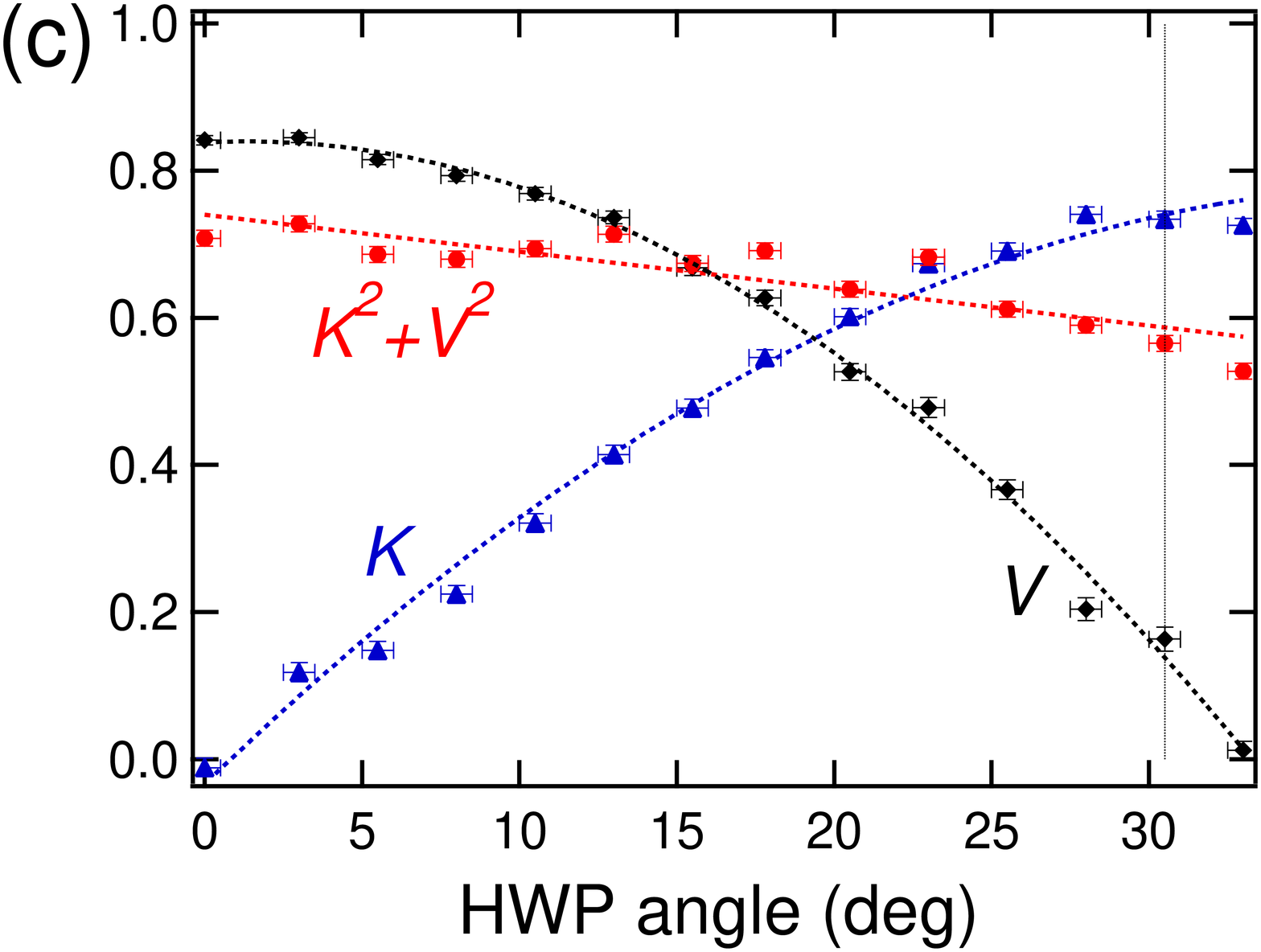}
\vspace{-0.6cm}
\caption{Non-destructive weak measurement of a polarisation encoded single photon qubit. (a) No measurement: The real part of output density matrix $\rho$ of the signal qubit for $\alpha=0$ with purity Tr$[\rho^2]$=0.89. (b) Strong measurement: The real part output density matrix of the signal qubit for $\alpha=\frac{\sqrt{3}}{2}$ with purity 0.51. (Note that all the imaginary components for \textbf{A} and \textbf{B} (not shown) are $<0.08$.) (c) Weak measurement: Plots of $K$, $V$ and $K^2+V^2$ for a number of different values of $\alpha$ ($\theta$ is the angle of the HWP used to set the polarisation of the meter and the dotted vertical line shows the expected position of $\alpha=\frac{\sqrt{3}}{2}$). The error bars take into account the count statistics and the input wave plate angle setting respectively.}
\label{weak}
\vspace{-0.7cm}
\end{center}
\end{figure}

Along with performing strong measurements of polarisation, our device also allows for non-destructive weak, measurements. We vary the input state of the meter $|\Psi\rangle_m=\alpha|H\rangle_m+\beta|V\rangle_m$ to vary the strength of the measurement. We also introduce a $\tfrac{2}{3}$ loss in the $s_V$ mode [as shown in Fig. \ref{schematic}(b)] to balance the measurement statistics. The most interesting behaviour can be seen for an equal superposition signal input, eg $(|H\rangle_s+|V\rangle_s)/\sqrt{2})$. The output state for an arbitrary meter input is then
\begin{eqnarray*}
|\phi_{out}\rangle&=&\tfrac{1}{2\sqrt{3}}[|H\rangle_s(\sqrt{\tfrac{2}{3}}\alpha|H\rangle_m+\sqrt{2}\beta|V\rangle_m)\\
&+&|V\rangle_s(\sqrt{\tfrac{2}{3}}\alpha|H\rangle_m+\sqrt{2}\beta|V\rangle_m)]+...
\end{eqnarray*}
where again the terms not shown are failure mechanisms. 

We can characterise this weak measurement by measuring the 1-qubit reduced density matrix of the signal output [Fig. \ref{weak}(a) and (b)]. The $H$ and $V$ populations do not change, regardless of the meter input state. However for $|\alpha|=0$ we observe a coherent superposition, while for $\alpha=\frac{\sqrt{3}}{2}$ we have an incoherent mixture as expected when a strong measurement is made. In the intermediate region the signal output is partially mixed. The degree of coherence can also be determined by measuring the visibility $V$ of the linear polarisation fringes as shown in Fig. \ref{weak}(c). These results explicitly demonstrate the \emph{decoherence} that would appear in quantum cryptography due to an eavesdropper using weak QND measurement of polarisation, as  simulated in \cite{na-prl-84-4733}.

The fundamental principle of complementarity \cite{bohm}, in particular wave-particle duality, can be tested with our general measurement in the fashion originally proposed in Ref. \cite{sa-pra-39-694}. In the spatial interferometer of Fig. \ref{exp}, $K$ quantifies the degree of ``which-path" information,  and $V$ the quantum indistinguishability. They must satisfy $V^2+K^2\leqslant1$ \cite{en-prl-77-2154}. In Fig. \ref{weak}(a) we plot $K$, $V$, and $K^2+V^2$ for a range of values of $\alpha$. As $K$ increases, $V$ decreases. Ideally our weak QND scheme is optimal:  $K^2+V^2=1$ for all meter polarizations. In our experiment $K^2+V^2<1$ due to non-ideal mode matching. The decline of $K^2+V^2$ with increasing $K$ can be attributed to the increasing requirement for non-classical (as well as classical) interference as the strength of the QND measurement is increased. In contradistinction to the non-destructive scheme presented here, previous tests of complementarity have relied on encoding which-path information onto a different degree of freedom of the interfering particles \cite{du-nat-395-33,sc-pra-60-4285}, so that which-path information is only obtained destructively, when the particles are measured.

In summary, we have proposed, demonstrated, and characterised a non-deterministic scheme for general measurement of a flying qubit. In addition, we have introduced the first set of fidelity measures to characterise the quality of any QND measurement. Because we are able to measure these fidelities directly and prepare all input states with high fidelity, we have demonstrated the most comprehensive characterisation of a QND measurement to date. We find that our device performs well against all three requirements of a QND measurement, with fidelities greater than 80\% for all measures and all input states. Operating in the weak regime, we have performed a non-destructive test of complementarity.
\\

\vspace{-0.3cm}
We thank J. S. Lundeen and G. J. Milburn for helpful discussions. This work was supported by the Australian government, the Australian Research Council, the US National Security Agency (NSA) and Advanced Research and Development Activity (ARDA) under Army Research Office (ARO) contract number DAAD 19-01-1-0651.


\begin{thebibliography}{26}
\bibitem[{qnd()}]{qndnote1}
\bibinfo{note}{The correlation function is defined $C_{MS} = \langle \hat
  {O}_{M} \hat {O}_{S} \rangle/(\langle \hat {O}_{m}^{2} \rangle \langle \hat
  {O}_{s}^{2} \rangle)^\frac{1}{2}$, where $\hat {O}$ is the QND observable:
  the Stokes operator $\hat{S1}$ for our measurement; the quadrature amplitude
  $\hat X$ for CV systems.}
\end{thebibliography}
\end{document}